\documentclass[twocolumn,showpacs,preprintnumbers,amsmath,amssymb,prl,superscriptaddress]{revtex4-2}
\usepackage{multirow}
\usepackage{amsfonts}
\usepackage[english]{babel}
\usepackage[T1]{fontenc}
\usepackage{mathrsfs}
\usepackage{graphicx}
\usepackage{dcolumn}
\usepackage{bm}
\usepackage{times}
\usepackage[colorlinks,bookmarks=true,citecolor=blue,linkcolor=red,urlcolor=blue]{hyperref}
\usepackage[tight, FIGTOPCAP, hang, raggedright, nooneline]{subfigure}
\begin{document}

	\title{Energy Gap in Weakly Disordered Fractional Quantum Hall Liquids: \\ Quantitative Comparison to GaAs Quantum Well Experiments at $\nu= 1/3$}
	
	\author{Yi-Han Zhou}	
	\affiliation{Zhejiang Institute of Modern Physics, Zhejiang University, Hangzhou 310058, China}
	\author{Zi-Ang Wang}
	\affiliation{Zhejiang Institute of Modern Physics, Zhejiang University, Hangzhou 310058, China}
	\author{Xin Wan}
	\email{xinwan@zju.edu.cn}
	\affiliation{Zhejiang Institute of Modern Physics, Zhejiang University, Hangzhou 310058, China}
	\affiliation{Zhejiang Key Laboratory of Micro-Nano Quantum Chips and Quantum Control, School of Physics, Zhejiang University, Hangzhou 310027, China}
	\author{Zhao Liu}
	\email{zhaol@zju.edu.cn}
	\affiliation{Zhejiang Institute of Modern Physics, Zhejiang University, Hangzhou 310058, China}
	\affiliation{Zhejiang Key Laboratory of Micro-Nano Quantum Chips and Quantum Control, School of Physics, Zhejiang University, Hangzhou 310027, China}
	
	\date{\today}
	
	\begin{abstract}
		Based on a recent experiment in high-quality GaAs quantum wells [Phys. Rev. Lett. 127, 056801 (2021)], we present a microscopic study of the energy gap in two-dimensional electron gases at filling factor $\nu=1/3$, explicitly incorporating both finite layer thickness and disorder effects. The finite layer thickness is modeled by solving the Poisson-Schrödinger equations for the experimental devices, yielding the electron wave functions in the perpendicular direction. Using these and the disorder energy extracted from the experiment, we estimate the charge gap and the mobility gap at $\nu=1/3$ in the weakly disordered lowest Landau level. Remarkably, both gaps show good quantitative agreement with the activation gap measured from the experiment in narrow quantum wells. Our results also indicate the potential need of incorporating higher subbands to make accurate theoretical predictions of the energy gap in wide quantum wells.
	\end{abstract}

	\maketitle
		{\it Introduction.} Rich many-body phenomena emerge in two-dimensional electron gases (2DEGs) subjected to strong perpendicular magnetic fields and cooled to ultra-low temperatures. Among these, the fractional quantum Hall effect (FQHE) stands out as one of the most celebrated examples, having attracted tremendous attention since its discovery over 40 years ago~\cite{Tsui(1982)}. The intrinsic topological order of FQHE states strikingly distinguishes them from traditional phases characterized by symmetry breaking. This unique feature renders certain FQHE states promising platforms for topological quantum computation~\cite{Nayak(2008)}.
		
		Electrons form an incompressible liquid when the FQHE occurs. This incompressibility is characterized by a nonzero charge gap $\Delta_c$, which measures the energy cost to create a far-separated quasiparticle-quasihole pair in the system. In experiments at very low temperatures, the charge gap manifests itself in the thermally activated longitudinal resistance $R_{xx}\propto e^{-\frac{\Delta_a}{2k_B T}}$, where $k_B$ is the Boltzmann constant and $T$ is the temperature. The activation gap $\Delta_a$ depends on many details of the system, including the finite layer thickness of the 2DEG, the disorder strength, Landau level (LL) mixing, etc.
		
		For a long time, there has been interest in quantitatively comparing the theoretically calculated and experimentally measured energy gap, particularly for the strongest and first observed FQHE at filling factor $\nu=1/3$~\cite{Haldane(1985),Girvin(1985),Boebinger(1985),Willett(1988),Du(1993),MelikAlaverdian(1995),KPark(1999),MorfRH(2002),WanXin(2005),Pan(2020),Villegas(2021)}. On the theoretical side, numerical simulations in the lowest Landau level (LLL) for an ideal 2DEG without disorder and layer thickness yield $\Delta_c\approx0.1e^2/(4\pi\epsilon l_{B})$~\cite{Haldane(1985),Girvin(1985)}, where $\epsilon$ is the dielectric constant and $l_{B}=\sqrt{\hbar/(eB)}$ is the magnetic length. Efforts have also been made to incorporate realistic factors, such as finite layer thickness~\cite{Zhang(1986),MelikAlaverdian(1995),KPark(1999),MorfRH(2002),Zhao(2022)} and LL mixing (LLM)~\cite{MelikAlaverdian(1995),Yoshioka(1984),Yoshioka(1986),Zhao(2022)}. Although numerical calculations successfully capture the suppression of the gap with increasing layer thickness and LLM, a recent experiment that systematically extracted the activation gap at $\nu=1/3$ in extremely high-quality 2DEGs reported appreciably lower $\Delta_a$ than theoretical $\Delta_c$ obtained in clean systems, even after the reduction of energy gap caused by disorder has been empirically compensated~\cite{Villegas(2021)}.
		
		In this work, we aim to resolve the discrepancy between the experimental gap data in Ref.~\cite{Villegas(2021)} and theoretical predictions. To accurately account for the finite thickness of realistic 2DEG samples, we consider a typical modulation-doped GaAs quantum well structure that approximates the experimental setup of Ref.~\cite{Villegas(2021)}. Instead of using trial wave functions~\cite{Fang(1966),Zhang(1986),WanXin(2005),MorfRH(2002)}, we numerically solve the self-consistent Poisson-Schr\"{o}dinger equations~\cite{Mooney(1940),von(1983),Sebawe(2016)} to extract the electron subband wave function in the perpendicular direction. On the other hand, while most previous theoretical works focused on clean systems, we quantitatively model the disorder in experimental devices instead of empirically compensating the disorder effect. In the presence of random impurities, excitations localized by the disorder cannot contribute to the longitudinal resistance. Therefore, the mobility gap $\Delta_m$, the energy required to excite delocalized quasiparticles and quasiholes, is a proper estimate of the experimentally measured $\Delta_a$~\cite{Sheng(2003),WanXin(2005)}. By contrast, the charge gap $\Delta_c$ may underestimate $\Delta_a$ due to the disorder induced localized excitations. However, we expect this difference to be sufficiently small in ultra-high quality samples with very weak disorder. To corroborate this point, we keep the lowest subband and compute $\Delta_c$ and $\Delta_m$ using disorder strengths extracted from the experimental data. Remarkably, they are indeed close to each other, and both show good quantitative agreement with experimentally measured $\Delta_a$ for various narrow quantum wells in Ref.~\cite{Villegas(2021)}. Since the computation of $\Delta_c$ only requires the lowest energies under periodic boundary conditions, our work provides a more efficient method for the quantitative comparison between theoretical and experimental data of the energy gap in weakly disordered samples. In wide quantum wells, our theoretical prediction of the energy gap shows discrepancy from the experimental data, which may result from the effects of higher subbands.
		
		{\it Model.} We consider $N_{e}$ spin-polarized electrons of charge $-e$ on an $L\times L=2\pi N_\phi l_{B}^2$ square torus with a perpendicular uniform magnetic field, where $N_\phi$ is the number of magnetic flux quanta penetrating the torus. The LL filling factor is defined as $\nu=N_{e}/N_{\phi}$. The Coulomb energy $E_{C}=e^2/(4\pi\epsilon l_{B})$ is used as the energy unit in this section. For numerical efficiency, we neglect the LLM which is suppressed at strong magnetic fields. The many-body Hamiltonian projected to the LLL, consisting of the interaction and disorder potential, takes the form
		\begin{eqnarray}
		\label{eq:totalH}
				H&=&\sum_{m_1,m_2,m_3,m_4=0}^{N_{\phi}-1}V_{m_1,m_2,m_3,m_4}c^{\dagger}_{m_1}c^{\dagger}_{m_2}c_{m_3}c_{m_4}\nonumber\\
				  &+&\sum_{m_1,m_2=0}^{N_{\phi}-1}U_{m_1,m_2}c^{\dagger}_{m_1}c_{m_2},
		\end{eqnarray}
		where $c^{\dagger}_{m}$ and $c_{m}$ are the operators creating and annihilating an electron in the LLL orbital $m$, respectively. We impose the twisted boundary condition $\mathcal{T}_{\mu}\psi_m({\bf r})=e^{i\theta_\mu}\psi_m({\bf r})$ on the LLL orbitals $\psi_m({\bf r})$, where $\mathcal{T}_\mu$ is the magnetic translation operator in the $\mu=x,y$ direction and $\theta_\mu$ is the boundary phase. Under the Landau gauge $\bm{A}=(0,-Bx)$, the interaction matrix element $V_{\{m_i\}}$ and disorder matrix element $U_{m_1,m_2}$ are
		\begin{eqnarray}
					V_{\{m_i\}}=\frac{1}{2}\delta^{\text{mod}\ N_{\phi} }_{m_1+m_2,m_3+m_4}\sum_{s,t=-\infty}^{+\infty}\delta^{\text{mod}\ N_{\phi}}_{t,m_1-m_4}V_{\bm{q}}\nonumber\\
				\times e^{-\frac{1}{2}q^2 l_B^2} e^{i\frac{2\pi s}{N_{\phi}}(m_1-m_3)} e^{i\frac{\theta_x}{N_\phi}(m_1+m_2-m_3-m_4)}
			\label{eq::Vm}
		\end{eqnarray}
		and 
		\begin{eqnarray}
				U_{m_1,m_2}=\sum_{s,t=-\infty}^{+\infty}\delta_{t,m_1-m_2}^{\text{mod}\ N_{\phi}}U_{\bm{q}}e^{-\frac{1}{4}q^2 l_B^2}\nonumber\\
				\times e^{i\frac{\pi s}{N_{\phi}}(2m_1-t+\theta_y/\pi)}e^{i\frac{\theta_x}{N_\phi}(m_1-m_2-t)},
		\end{eqnarray}
		respectively, where $\delta_{i,j}^{\text{mod}\ N_{\phi}}$ is the periodic Kronecker delta function with period $N_{\phi}$, $\bm{q}=(q_x,q_y)=(\frac{2\pi s}{L},\frac{2\pi t}{L})$ with integers $s$ and $t$, $q\equiv |{\bm q}|$, and $V_{\bm q}$ and $U_{\bm q}$ are the Fourier transforms of the interaction potential and the impurity potential, respectively. We use the Coulomb interaction with $V_{\bm{q}}=\frac{1}{N_{\phi}}\frac{1}{ql_B}F(q)$, where the factor $F(q)$ accounts for the finite layer thickness and depends on the electron distribution along the $z$-axis (the form will be given later). The $s=t=0$ must be excluded from the sum in Eq.~(\ref{eq::Vm}) to remove the $q=0$ singularity of $V_{\bm{q}}$. Physically this can be achieved by involving the interaction of electrons with their images and a neutralizing background charge~\cite{PhysRevB.29.6833}. We use the Gaussian white noise to model the disorder potential $U({\bm r})$, which satisfies $\langle U({\bm r})\rangle=0,\langle U({\bm r})U({\bm r}')\rangle=W^2l_B^2 \delta({\bm r}-{\bm r}')$. The disorder strength is measured by $W$, and $\langle\cdots\rangle$ represents the average over disorder configurations. Accordingly, we have
		\begin{equation}
			\langle U_{\bm q}\rangle=0, \ \langle U_{\bm{q}}U_{\bm{q}^{\prime}}\rangle=\frac{W^2}{2\pi N_{\phi}}\delta_{\bm{q},-\bm{q}^{\prime}}
			\label{eq::GW}
		\end{equation}
		in the momentum space. In each disorder configuration, the real $U_{{\bm q}={\bm 0}}$ is generated from a Gaussian distribution with zero mean and variance $\frac{W^2}{2\pi N_\phi}$. By contrast, $U_{{\bm q}\neq{\bm 0}}$ is complex, whose real part and imaginary part are separately produced from a Gaussian distribution with zero mean and variance $\frac{W^2}{4\pi N_\phi}$. The Hermiticity of the Hamiltonian requires $U_{{\bm q}}^*=U_{-{\bm q}}$, so we only generate independent $U_{{\bm q}}$'s with ${\bm q}\in\{{\bm q}|q_x=0,q_y\geq 0\}\cup\{{\bm q}|q_x>0\}$.
		
\begin{figure}
			\centerline{\includegraphics[width=\linewidth]{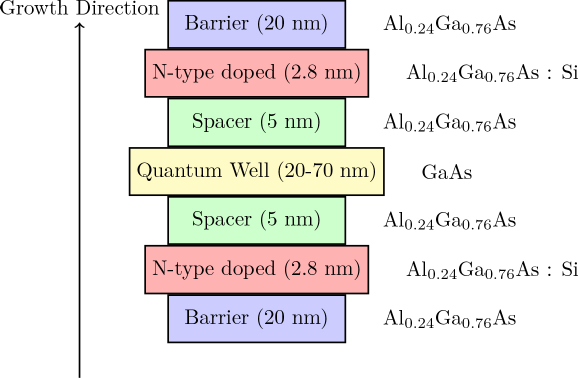}}
			\caption{Schematic of the modulation-doped structure that we use in the Poisson-Schr\"{o}dinger solver, which is symmetric with n-type doping layers on both sides of the GaAs quantum well, separated by spacer layers.}
\label{fig:heterostructure}
		\end{figure}

		{\it Poisson-Schr\"{o}dinger equations.} The experiment in Ref.~\cite{Villegas(2021)} utilizes the doping well structure~\cite{dopingwell} to achieve extremely high-mobility 2DEGs. For numerical convenience, we employ the standard symmetric modulation-doped geometry, which is a robust computational model suitable for Poisson-Schr\"{o}dinger solvers~\cite{Tan(1990), Mohan(2004)}, to capture the finite thickness of 2DEGs in the realistic experimental devices. The modulation-doped structure simulated by us is displayed in Fig.~\ref{fig:heterostructure}, which is symmetric with an Al\(_{0.24}\)Ga\(_{0.76}\)As barrier (the same as in Ref.~\cite{Villegas(2021)}), an n-type doping layer, and a spacer on both sides of the GaAs quantum well. The layer thicknesses $w$ of the quantum well varies from $20 \ {\rm nm}$ to $70 \ {\rm nm}$. Although the quantum wells are flanked by 150-nm-thick Al\(_{0.24}\)Ga\(_{0.76}\) barriers in Ref.~\cite{Villegas(2021)}, we choose a reduced barrier thickness (e.g., $20 \ {\rm nm}$) which is verified to be sufficient for the convergence of our Poisson-Schr\"{o}dinger solver. The donor concentration $N_d^+$ in the simulation is calibrated to strictly reproduce the experimental planar electron density $n_s \approx 1.1 \times 10^{11} \text{cm}^{-2}$, ensuring the electronic states in the quantum well are physically comparable to the experiment.
		
		For the 2DEG in such a multi-layer structure, we assume separability of the planar and perpendicular degrees of freedom. We use the model of two-dimensional free electrons to describe the degree of freedom in the $xy$ plane. The motion of electrons in the vertical $z$-direction depends on the subband wave functions $\eta_i(z)$, which are governed by Schrödinger and Poisson equations~\cite{BenDaniel(1966),Stern(1984),Ortalano(1997),jain(2007)}
		\begin{eqnarray}
				&&E_i\eta_i(z)=\left[\frac{1}{4}\left(M^{\eta}\hat{\bm{P}}M^{\xi}\hat{\bm{P}}M^{\sigma}+{\rm h. c.}\right)  +V(z)\right]\eta_i(z), \nonumber\\
				&&\frac{\partial}{\partial z}\left(\epsilon(z)\frac{\partial V_{p}(z)}{\partial z}\right)=\rho(z).
		\end{eqnarray}
		Here $E_i$ are the subband energies, and $\hat{\bm{P}}=-i\hbar\partial/ \partial z$ is the momentum operator. The dielectric function \(\varepsilon(z)\) and effective mass \(M(z)\) are piecewise-constant within each layer, with discontinuities only at heterointerfaces. In general, $\eta+\sigma+\xi=-1$ is required to ensure the correct dimension of the Hamiltonian. Following Refs.~\cite{Hebal(2021),Sebawe(2016)}, we set $\eta=\sigma=0$ in our calculation. The total potential energy experienced by electrons is
		\begin{equation} 
		\label{eq:vz}
		V(z)=V_0(z) + V_p(z),
		\end{equation} 
		where $V_0(z)$ is the band offset at the heterointerfaces, and we leave out the exchange-correlation potential correction~\cite{Ortalano(1997),KPark(1999),MorfRH(2002)} to simplify our model. The term $V_p(z)$ originates from the $z$-direction charge distribution 
		\begin{equation}
		\label{eq:rhoz}
		\rho(z) = -e \left[ N_d^+(z) - n(z) \right],
		\end{equation}
		where \(N_d^+(z)\) is the volume density of ionized donors, chosen as a step function which is uniform in the two doped layers. \(n(z)\) is the volume density of electrons obtained from $\eta_i(z)$ via Fermi–Dirac statistics~\cite{Harrison(2016)}:
		\begin{equation}
		\label{eq:nz}
			n(z)=\frac{m^* k_{B}T}{\pi\hbar^2 A}\sum_i\ln\left(1+e^{\frac{\mu-E_{i}}{k_{B}T}}\right)|\eta_{i}(z)|^2,
		\end{equation}
		where $m^*$ is the electron's effective mass in the GaAs layer, $T$ is the temperature, $\mu$ is the electron's chemical potential, and $A$ is the area of the 2DEG. 
		
		The Poisson and Schrödinger equations are coupled with each other by Eqs.~(\ref{eq:vz}), (\ref{eq:rhoz}), and (\ref{eq:nz}). We employ a finite-difference scheme to self-consistently solve the coupled Poisson-Schrödinger equations, working in the zero-temperature limit. Initially we choose $V(z)=V_0(z)$ and solve the Schr\"{o}dinger equation to obtain $E_i$ and $\eta_i(z)$. Then we use Eqs.~(\ref{eq:rhoz}) and (\ref{eq:nz}) to compute $\rho(z)$. With $\rho(z)$, $V_p(z)$ can be extracted by solving the Poisson equation. After adding $V_{p}(z)$ into $V(z)$, we solve Schr\"{o}dinger equation again to update $E_i$ and $\eta_i(z)$. We repeat these procedures until convergence. We also properly adjust $\mu$ to make sure the total planar density of electrons in the quantum well converges to $1.1\times 10^{11}\text{ cm}^{-2}$ as achieved in Ref.~\cite{Villegas(2021)}.
		
\begin{figure}
			\centerline{\includegraphics[width=\linewidth]{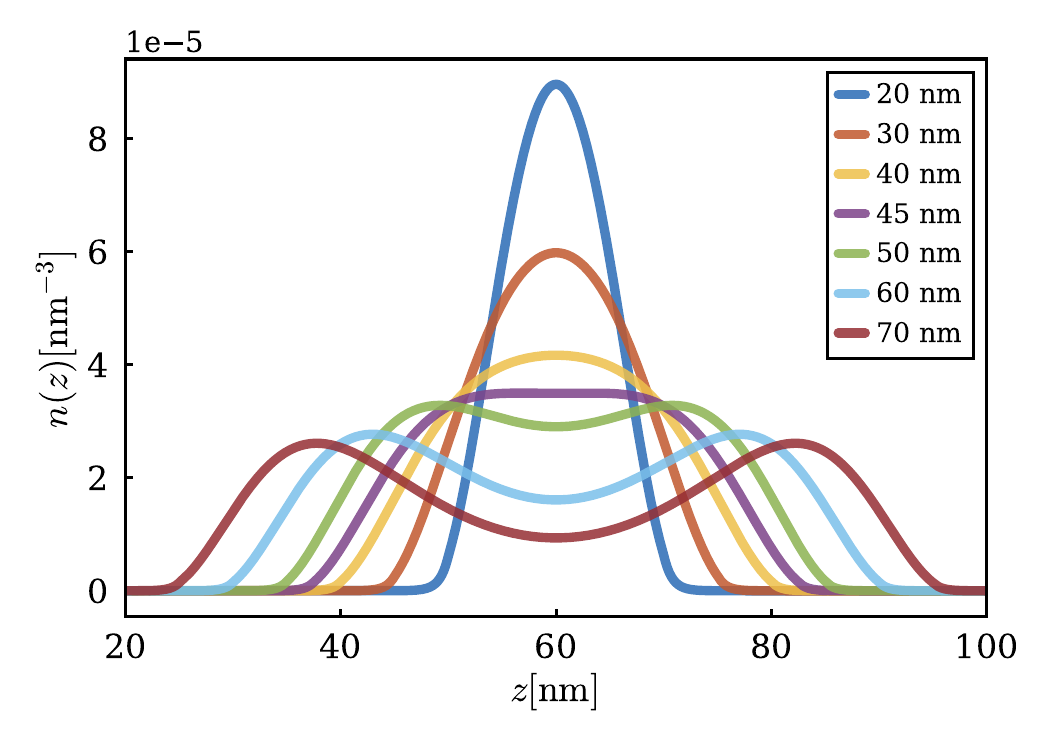}}
			\caption{The charge distribution $n(z)$ in quantum wells. The layer thickness $w$ varies from $20 \ {\rm nm}$ to $70 \ {\rm nm}$.}
			\label{F_charge_distribution}
		\end{figure}

		In Fig.~\ref{F_charge_distribution}, we present the charge distribution $n(z)$ in the quantum well for different layer thickness $w$. When $w$ is relatively small, $n(z)$ has a single peak whose width grows with increasing $w$. However, $n(z)$ develops two peaks when $w$ reaches $50 \ {\rm nm}$, corresponding to a wide quantum well. 
The authors of Ref.~\cite{Villegas(2021)} also simulated the charge distribution in the quantum well by their own Poisson-Schr\"odinger solver. Notably, our results faithfully reproduce the evolution of charge distribution reported in Ref.~\cite{Villegas(2021)}, capturing the critical transition of $n(z)$ from a single-peak to a bilayer-like distribution as the well width increases beyond $50 \ {\rm nm}$. 
		
		{\it Charge gap.} With the $z$-direction subband wave functions evaluated for the 2DEGs prepared in Ref.~\cite{Villegas(2021)}, we now compute the charge gap using the microscopic model Eq.~(\ref{eq:totalH}). For numerical efficiency, we only keep the lowest subband, so that 
		\begin{equation}
			F(q)= \int_{-\infty}^{+\infty}\!dz_1\int_{-\infty}^{+\infty}\!dz_2 |\eta_0(z_1)|^2 |\eta_0(z_2)|^2 e^{-q|z_1-z_2|}.
		\end{equation}
		Generally, $F(q)$ softens the Coulomb interaction at short distances. 
		In experiments, fitting the activation gap at $\nu=n/(2n+1)$ into $\Delta_a=\frac{1}{2n+1}E_0-\Gamma$ is often used to estimate the disorder energy $\Gamma$ in the 2DEG. $\Gamma$ can also be interpreted as the impurity induced broadening of the $\Lambda$ levels of composite fermions (CFs)~\cite{Du(1993),jain(2007)}. Therefore, we choose the $\Gamma$ extracted in Ref.~\cite{Villegas(2021)} as the strength $W$ of the Gaussian white noise in our microscopic model [see the Supplemental Materials (SM) for details], which we list in Tab.~\ref{T_thickness_and_W} for different quantum well widths. 
		\begin{table}
			\caption{Widths $w$ and disorder energies $\Gamma$ of the quantum wells fabricated in Ref.~\cite{Villegas(2021)}. $E_C$ is the Coulomb energy determined by the magnetic length at $\nu=1/3$, which is $l_{B}=7.1\text{ nm}$ in Ref.~\cite{Villegas(2021)}.}
			\label{T_thickness_and_W}
			\begin{ruledtabular}
				\begin{tabular}{cccccccc}
					$w$ (nm) & $20$ & $30$ & $40$ & $45$ & $50$& $60$& $70$ \\
					\hline
					$\Gamma$ $(E_C)$ & $0.014$ & $0.005$ &$0.007$& $0.008$& $0.018$ & $0.007$ & $0.007$ \\
				\end{tabular}
			\end{ruledtabular}
		\end{table}			
		
		We first evaluate the charge gap at $\nu=1/3$ for a specific number of electrons under periodic boundary conditions using the expression~\cite{MorfRH(2002),WanXin(2005)} (see the SM for a detailed derivation) 		
		\begin{eqnarray}
				\Delta_c(N_e)&=&E_g(N_e,N_{\phi,0}+1)+E_g(N_e,N_{\phi,0}-1)\nonumber\\
				&-&2E_g(N_e,N_{\phi,0})-\frac{2}{9}S,
				\label{eq::chargegap}
		\end{eqnarray}  
		where $E_{g}(N_{e},N_\phi)$ is the lowest energy obtained by exact diagonalization (ED) for the Hamiltonian Eq.~(\ref{eq:totalH}) with $N_{e}$ electrons and $N_\phi$ magnetic flux quanta. The first two terms in Eq.~(\ref{eq::chargegap}) are the lowest energies when a single quasihole and a single quasiparticle is excited, respectively, and the third term with $N_\phi=N_{\phi,0}\equiv 3N_e$ corresponds to the absence of quasiparticles and quasiholes. The last term $-2S/9$ accounts for the interaction of quasihole/quasiparticle with its periodic images on the torus geometry~\cite{Haldane(1985),WanXin(2005)}, where $S$ is the classical energy per electron for a Wigner crystal with the square $L\times L$ unit cell~\cite{Bonsall(1977),Fujiki(1992)}. To create a single quasihole (quasiparticle), we add (remove) one magnetic flux quantum with respect to $N_{\phi,0}$, which is achieved by varying the magnetic field (equivalently, the magnetic length) at fixed length of the square torus. We choose the Coulomb energy $E_C$ determined by the magnetic length at exactly $\nu=1/3$ as the energy unit. In the experiment Ref.~\cite{Villegas(2021)}, this magnetic length is $l_{B}=7.1\text{ nm}$. For the energies at other magnetic fields, we convert their values in units of $E_C$ using $E_C'=\sqrt{N_{\phi}'/N_{\phi,0}}E_C$, where $N_\phi'=N_{\phi,0}\pm 1$. 
For disordered systems, we also need to average $E_{g}(N_{e},N_{\phi,0}\pm 1)$ and $E_{g}(N_{e},N_{\phi,0})$ over various impurity configurations. The disorder strength $W$ in Eq.~(\ref{eq::GW}) does not change with the magnetic field and $N_\phi$.
	
\begin{figure}
			\centerline{\includegraphics[width=\linewidth]{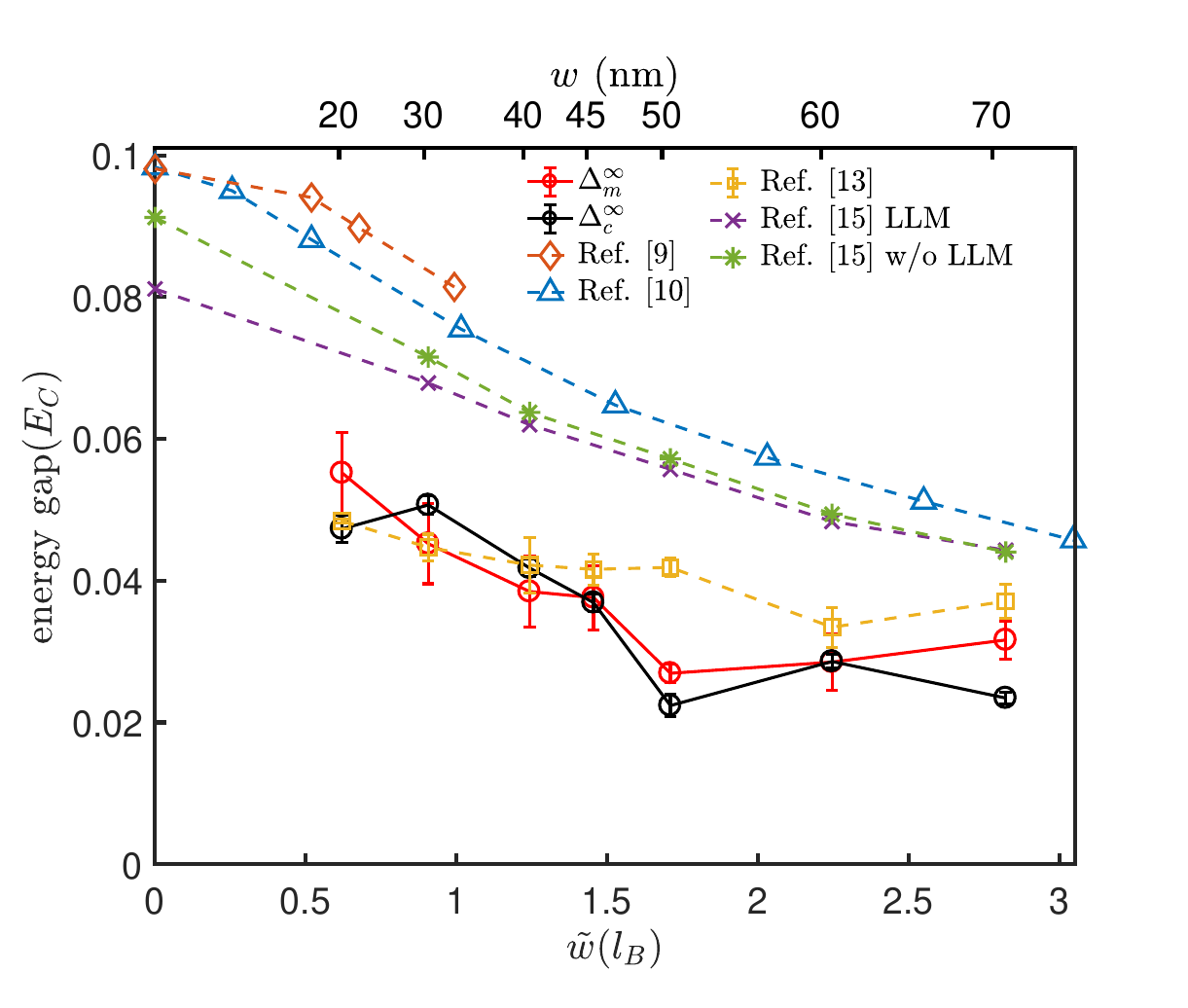}}
			\caption{Estimations of the charge gap $\Delta^{\infty}_c$ (black circles) and the mobility gap $\Delta_m^{\infty}$ (red circles) in the thermodynamic limit.  
			For the charge gap, we consider 800 disorder configurations for $N_e=5-8$ and 400 configurations for $N_e=9$. The finite-size data with $N_e=5-9$ are used in the extrapolation to the thermodynamic limit.
			For the mobility gap, we consider 200 configurations for $N_e=5-8$. We use the finite-size data with $N_e=5-8$ in the extrapolation to the thermodynamic limit except at $w=50\ {\rm nm}$, where we additionally compute 50 configurations for $N_e=9$ and use the data with $N_e=6-9$ in the extrapolation. The results of Ref.~\cite{Zhao(2022)} obtained with and without (w/o) the LLM are both presented. }
			\label{F_gap_vs_thickness}
		\end{figure}

		We estimate the charge gap $\Delta_{c}^{\infty}$ in the thermodynamic limit by fitting the finite-size data of $N_e=5-9$ electrons into $\Delta_{c}(N_{e})=\Delta_{c}^{\infty}+a/N_{e}$. In Fig.~\ref{F_gap_vs_thickness}, we present $\Delta_{c}^{\infty}$ as a function of effective layer thickness $\tilde{w}$ (the standard deviation of the charge distribution as defined in Ref.~\cite{Villegas(2021)}) as well as the actual quantum well width $w$. We also present in the same figure the charge gaps reported in other theoretical studies for clean systems~\cite{KPark(1999),MorfRH(2002),Zhao(2022)}. Ref.~\cite{KPark(1999)} utilized the Poisson-Schr\"{o}dinger equations to capture the finite thickness, and relied on variational Monte Carlo (VMC) simulations of CF trial wave functions to extract the charge gap. Ref.~\cite{MorfRH(2002)} accounted for the finite thickness by using the Gaussian and $z\times$Gaussian trial wave functions for the lowest subband, and the charge gap was computed by ED on the sphere geometry. Ref.~\cite{Zhao(2022)} treated the finite layer thickness using the Poisson-Schr\"{o}dinger equations on the spherical geometry, and the charge gap was calculated by VMC and ED either with or without the LLM. Our work, like these studies, captures the decaying trend of $\Delta_{c}$ with increasing layer thickness. However, our data fall below other theoretical studies due to inclusion of disorder.

		{\it Mobility gap.} The charge gap represents the energy cost of creating a far-separated quasiparticle-quasihole pair, regardless of being localized or not. In the mean-field level, because the $\Lambda$ levels of CFs are broadened by disorder, the mobility gap, which measures the energy cost of exciting delocalized quasiparticles and quasiholes, should be larger than the charge gap (see the SM). However, in this work we consider extremely high-quality samples in which the disorder strengths (Tab.~\ref{T_thickness_and_W}) are much weaker than the critical value required to destroy the Laughlin topological order~\cite{WanXin(2005)}, so we expect small difference between the charge gap and the mobility gap. 
		
		To corroborate this point, we explicitly evaluate the mobility gap at $\nu=1/3$ using the same subband wavefunctions and disorder strengths as in the previous section. Following the method proposed in Refs.~\cite{Sheng(2003),WanXin(2005)}, we use the many-body Chern number~\cite{Niu(1985),Fukui(2005)} of the system's eigenstates to probe the mobility gap. Due to the three-fold degeneracy of each energy level in the thermodynamic limit for the $\nu=1/3$ FQH liquid on the torus, we consider the Chern numbers of groups of states for each system size $N_e$, with the $n_{g}$-th group containing the $(3n_{g}-2)$-th, $(3n_{g}-1)$-th, and $(3n_{g})$-th eigenstates (sorted in ascending order of energy) of Eq.~(\ref{eq:totalH}). These eigenstates are obtained by ED for a fixed disorder configuration $U_{\bm q}$. We then twist the boundary conditions without changing the disorder configuration to get the total Chern number of the group. The details of the many-body Chern number calculation are provided in the SM. 
		
		Repeating the procedures above for various disorder configurations, we can evaluate the probability distribution $P(C)$ of the total Chern number $C$ for a specific group of states. We find that the FQH ground-state manifold ($n_g=1$) always satisfies the condition $P(C=1)=1$. Thus, the Chern number per state is $1/3$, matching the quantized Hall conductance of the $\nu=1/3$ Laughlin state. For groups beyond the ground-state manifold, we notice that $P(C)$ starts to develop weights on $C\neq 1$ due to disorder. If $P(C=1)$ remains close to $1$ for a given group, the excitations in that group can be regarded as localized quasiparticle-quasihole pairs which do not affect the Hall resistance or the longitudinal resistance. On the contrary, the sharpest reduction of $P(C=1)$ is a striking signal of delocalized excitations that can significantly change the transport of the system, therefore we identify the corresponding group as the mobility edge. The mobility gap for a given system size is then estimated as the disorder-averaged difference between the mean energy of the group at the mobility edge and the ground-state group (see the SM for a concrete example). 
		
		Unlike calculating the charge gap, which requires only the lowest energy under periodic boundary conditions, estimating the mobility gap involves several groups of excited states and twisted boundary conditions.  As a result, for most values of $w$, we deal with at most 
$N=8$ electrons, and the number of disorder configurations is reduced to one quarter of that used for calculating the charge gap.

		Having estimated the mobility gap $\Delta_{m}(N_{e})$ for each system size, we extrapolate it to the thermodynamic limit by fitting the finite-size data into $\Delta_{m}(N_{e})=\Delta_{m}^{\infty}+b/N_{e}$. As shown in Fig.~\ref{F_gap_vs_thickness}, $\Delta_{m}^{\infty}$ is indeed closely followed by the charge gap $\Delta_{c}^{\infty}$. While the CF picture suggests $\Delta_{m}^{\infty}<\Delta_{c}^{\infty}$, we find that in our estimation the mobility gap lies below the charge gap at the quantum well widths $w=30\ {\rm nm}$ and $w=40\ {\rm nm}$. We attribute this discrepancy to the larger uncertainties in estimating the mobility gap, arising from the limited number of disorder configurations and the finite system sizes.
				
		 {\it Comparison with the activation gap.} We now compare our theoretical gap estimations with the experimental data from Ref.~\cite{Villegas(2021)}. Overall, both the theoretical charge gap and mobility gap are consistent with the activation gaps reported in transport measurements, as shown in Fig.~\ref{F_gap_vs_thickness}. In particular, they exhibit good quantitative agreement for narrow quantum wells. Compared with previous theoretical works in clean systems that also considered finite thickness via ansatz wavefunctions or Poisson-Schr\"odinger solver, the significantly suppressed discrepancy between the theoretical prediction in our work and the experimentally measured gap demonstrates the importance of including disorder in numerical simulations. Given that the computational cost of calculating the charge gap is significantly lower than that of the mobility gap, our findings suggest that the charge gap serves as a reliable and more computationally efficient proxy for estimating the activation gap in weakly disordered systems.
		 		 
		 We also notice discrepancies between our theoretical gaps and the experimental data for wide quantum wells at $w\geq 50 \ {\rm nm}$. This may be attributed to the reduced energy spacing between two adjacent subbands, making the approximation of keeping only the lowest subband less accurate at large $w$. Indeed, our Poisson-Schr\"odinger solver reveals appreciable occupation of electrons in higher subbands when $w\geq 50$ nm, consistent with the findings of Ref.~\cite{Villegas(2021)}. Moreover, at $w = 50\ {\rm nm}$, the electron charge distribution $n(z)$ undergoes a transition from a single-peak to a double-peak profile, marking the crossover from a single-layer 2DEG to a quasi-bilayer system, as illustrated in Fig.~\ref{F_charge_distribution}. Ref.~\cite{Villegas(2021)} reported that the Laughlin phase is replaced by a bilayer Wigner crystal state when $w>70\ {\rm nm}$. This weakening of the Laughlin phase in wide quantum wells may lead to larger finite size effects, thereby undermining the agreement between ED calculations and experimental data.
		 				
		{\it Conclusion and discussion.} In this work, we combine a Poisson–Schrödinger solver with ED to simulate the recently measured extremely high-mobility 2DEG samples. By explicitly incorporating both the finite quantum-well thickness and weak impurity potentials, we perform a direct comparison between theoretical estimates of the charge and mobility gap for the $\nu=1/3$ fractional quantum Hall state and the experimental activation gap extracted from the Arrhenius behavior of the longitudinal resistance. The theoretical gaps are consistent with the experimental values across a range of quantum-well widths, with particularly good quantitative agreement achieved in narrow quantum wells. Crucially, unlike the mobility gap, the charge gap can be computed without the need for excited states or twisted boundary conditions, making it a numerically far more accessible quantity for accurate theory-experiment comparison in weakly disordered systems. Our calculations can be straightforwardly applied to study the gaps at other filling factors, like $\nu=2/5$, $3/7$ and $5/2$.
		
		LLM is neglected in our calculations. However, it has been found in clean systems that LLM only causes small reductions of the energy gap for the quantum well widths considered in this work~\cite{Zhao(2022)} (see Fig.~\ref{F_gap_vs_thickness}). Therefore, we do not expect any substantial changes of our results even if LLM is included. 
		
		The significantly improved estimation of the energy gap in our work underscores the important role of disorder in bridging theoretical predictions and experimental measurements. While the disorder energy $\Gamma$ extracted in experiments might not be solely attributed to disorder~\cite{Zhao(2022)}, our results suggest it as a reasonable estimation of the disorder strength needed in numerical simulations. On the other hand, the discrepancy between our results and experimental data for wide quantum wells points to the necessity of incorporating additional subbands to achieve accurate quantitative prediction of the energy gap. It is worthy of investigating how this subband mixing affects the energy gap in wide quantum wells within and beyond the perturbative regime.
		
		{\it Acknowledgments.} This project was supported by the National Key Research and Development Program of China (Grant No. 2021YFA1401902). 
		
\bibliography{BibFQH}

\onecolumngrid

\appendix 
		\section{Appendix}
		\renewcommand{\theequation}{S\arabic{equation}}  
		\renewcommand{\thefigure}{S\arabic{figure}}  
		\renewcommand{\thetable}{S\arabic{table}} 
\setcounter{equation}{0} 
\setcounter{figure}{0} 
\setcounter{table}{0}  

	\subsection{A. Disorder strength}
\begin{figure}[htbp]
		\centerline{\includegraphics[width=0.4\linewidth]{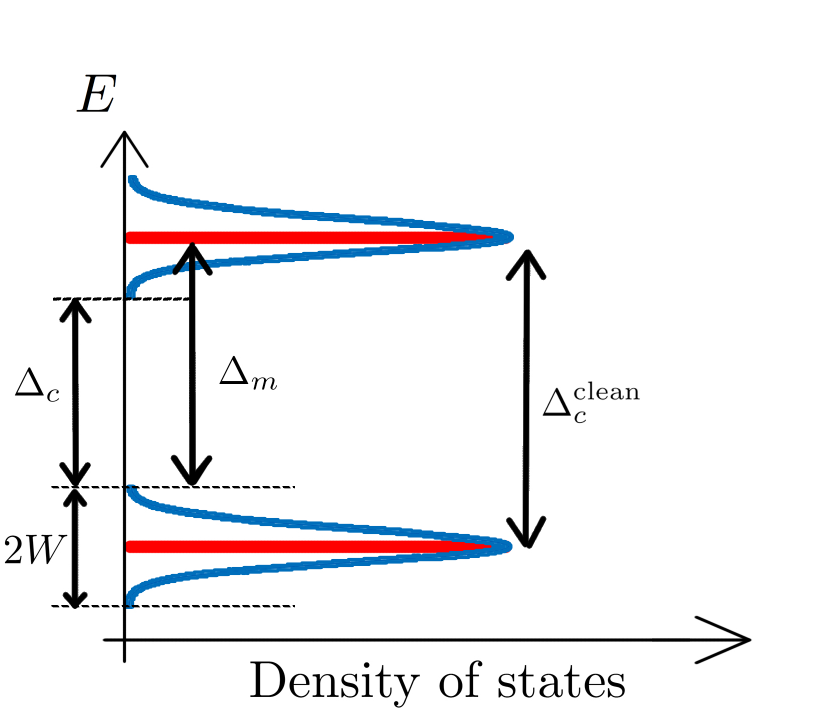}}
		\caption{Schematic illustration of the energy gap in an FQHE system. The horizontal and vertical axes represent the density of states and energy, respectively. In the CF picture, the $\nu=1/3$ state corresponds to $\nu^*=1$, i.e., the lowest $\Lambda$ level of CFs is fully filled. $\Lambda$ levels of CFs in the absence of disorder are indicated by the red lines, and their spacing is the charge gap $\Delta_{c}^{\rm clean}$ of the clean system. Disorder of strength $W$ broadens $\Lambda$ levels, as indicated by the blue lines. In the CF picture, we expect $\Delta_c^{\rm clean} - \Delta_{m}$, or equivalently,  $\Delta_c^{\rm clean}- \Delta_{a}$, equals $W$, where $\Delta_{m}$ and $\Delta_{a}$ are the mobility gap and the activation gap of disordered systems, respectively. Furthermore, we expect $\Delta_m- \Delta_c=W$, where $\Delta_c$ is the charge gap in a disordered system.}
		\label{F_disorder_energy}
	\end{figure}

	The disorder effect manifests itself in the reduced energy gap compared to that in the absence of disorder. Such a reduction can be understood using the composite fermion (CF) picture, where an electron is converted to a CF via attachment of $2p$ magnetic fluxes. At the mean-field level, CFs occupy their own Landau levels (LLs), dubbed $\Lambda$ levels, and the fractional quantum Hall effect (FQHE) of electrons at $\nu=n/(2pn+1)$ is interpreted as the integer quantum Hall effect of CFs at $\nu^*=n$~\cite{jain(2007)} ($n=p=1$ for $\nu=1/3$). Then the charge gap of the clean FQHE state, $\Delta_c^{\rm clean}$, is the energy spacing between two adjacent $\Lambda$ levels, as shown in Fig.~\ref{F_disorder_energy}. Like LLs of electrons, $\Lambda$ levels of CFs are broadened by disorder, and all states except those at the band centers are localized. 
	Given the strength $W$ of the Gaussian white noise, we expect the typical width of a $\Lambda$ band to be $\sim 2W$. In this case, the mobility gap $\Delta_m$, which is the energy cost to excite delocalized quasiparticles and quasiholes and directly corresponds to the activation gap $\Delta_a$ measured in experiments, is reduced by $W$ in a disordered system compared to the charge gap in the absence of disorder (Fig.~\ref{F_disorder_energy}). On the other hand, the charge gap in a disordered system, as shown in Fig.~\ref{F_disorder_energy}, is expected to be smaller than the mobility gap by $W$.
	
	For clean systems at $\nu=n/(2n+1)$, the charge gap scales as $\Delta_c^{\rm clean}=E_0/(2n+1)$, with $E_0\approx 0.3E_C$~\cite{Halperin(1993)}. Therefore we expect that the mobility and activation gaps tend to $-W$ with decreasing $1/(2n+1)$. The negative intercept of the activation gap, dubbed the disorder energy, was experimentally extracted in Ref.~\cite{Villegas(2021)} at various layer thicknesses. We adopt this disorder energy as the strength $W$ of Gaussian white noise in our microscopic model.
	
	\subsection{B. Expression of charge gap}
In this section, we derive the expression of the charge gap on the torus used in the main text. We adopt a simple heterojunction model consisting of a two-dimensional electron layer and a background layer. The electron layer has $N_e$ electrons and is penetrated by $N_\phi$ magnetic flux quanta. We assume a uniform positive charge distribution in the background layer, which is adjusted to possess a total charge of $+(N_e + Q)e$ with $e>0$. For filling factor $\nu=1/3$, $Q=0$, $Q=-1/3$ and $Q=1/3$ apply to the Laughlin ground state, a single quasiparticle excitation, and a single quasihole excitation, respectively, with $N_\phi=N_{\phi,0}\equiv 3N_e$, $N_\phi=N_{\phi,0}-1$, and $N_\phi=N_{\phi,0}+1$. For the ground state, the uniform positive background charge neutralizes the entire electron layer; otherwise, only the electron charge far from the quasiparticle/quasihole excitation is neutralized~\cite{Haldane(1985),WanXin(2005)}.
	
The total energy of the electron-background system can be decomposed into several contributions:
	\begin{equation}
		E=E_{\rm bb}+E_{\rm eb}+E_{\rm ee}+E_{\rm im},
	\end{equation}
where $E_{\rm bb}$ and $E_{\rm eb}$ represent the background-background and electron-background interaction energies, respectively, $E_{\rm ee}$ represent the energy of the electron layer including the electron-electron interaction and disorder, and $E_{\rm im}$ accounts for the interaction energy of an electron with its own periodic images. For convenience of notation, we assume zero distance between the electron and background layers and vanishing thickness for both layers.	In this case, we have
	\begin{equation}
		\begin{split}
			&E_{\rm bb}=\frac{(N_e+Q)^2}{2}\lim_{\bm{q}\to0}V_{\bm{q}},\\
			&E_{\rm eb}=-N_e(N_e+Q)\lim_{\bm{q}\to0}V_{\bm{q}},\\
			&E_{\rm ee}=\frac{N_e(N_e-1)}{2}\lim_{\bm{q}\to0}V_{\bm{q}}+E_g(N_e,N_\phi),\\
			&E_{\rm im}=\frac{N_e}{2}\lim_{\bm{r}\to0}(V(\bm{r})-v(\bm{r}))=\frac{N_e}{2}\lim_{\bm{q}\to0}(V_{\bm{q}}+2S).
		\end{split}
	\end{equation}
	Here $V_{\bm{q}}$ is the Fourier transform of the bare Coulomb potential $v(\bm{r})=\frac{e^2}{4\pi\epsilon |{\bm r}|}$,  $V(\bm{r})=\sum_{l_1,l_2}v(\bm{r}+l_1\bm{L}_1+l_2\bm{L}_2)$ is the periodic Coulomb potential on the torus spanned by $\bm{L}_1$ and $\bm{L}_2$~\cite{Fujiki(1992)}, $E_g(N_e,N_\phi)$ is the (non-singular) lowest energy of the electron's many-body Hamiltonian given in Eq.~(1) of the main text, and $S$ is the classical energy per electron (Madelung energy)~\cite{Bonsall(1977),PhysRevB.29.6833,WanXin(2005)} of a Wigner crystal of electrons with the torus as a unit cell. Combining all four terms above, we find the total energy as
	\begin{equation}
		E=\frac{Q^2}{2}\lim_{\bm{q}\to0}V_{\bm{q}}+N_e S+E_g(N_e,N_\phi).
		\label{eq::S3}
	\end{equation}
	
	Because $V_{\bm{q}}$ behaves as $1/q$, it has a singularity as $q\to 0$. If $Q=0$, the $q=0$ singularity in Eq.~(\ref{eq::S3}) vanishes, giving the total energy as $E(N_e,N_{\phi,0})=N_e S(L)+E_g(N_e,N_{\phi,0})$, where $S$ depends on the torus length $L=\sqrt{2\pi N_{\phi}}l_B$ ($l_B$ is the magnetic length). However, when a single quasiparticle or quasihole ($Q=\mp e/3$) is present, it is necessary to further subtract from Eq.~(\ref{eq::S3}) the interaction energy $E_{\rm im}^{\rm qp/qh}$ between the point-like quasiparticle or quasihole and its periodic images to remove the $q=0$ singularity~\cite{WanXin(2005)}. $E_{\rm im}^{\rm qp/qh}$ takes the form of
	\begin{equation}
		E_{\rm im}^{\rm qp/qh}=\frac{1}{2}\frac{(Qe)^2}{e^2}\lim_{\bm{r}\to0}(V(\bm{r})-v(\bm{r}))=\frac{Q^2}{2}\lim_{\bm{q}\to 0}V_{\bm{q}}+Q^2S,
	\end{equation}
	 	where the last equality again uses the definition of the Madelung energy $S$. Consequently, the total energy of the system with a single quasiparticle/quasihole becomes $E(N_e,N_\phi=N_{\phi,0}\mp 1)=E_g(N_e,N_\phi=N_{\phi,0}\mp 1)+N_eS(L)-S(L)/9$. In our calculation, we vary $N_{\phi}$ by changing the magnetic strength $B$ while keeping the physical dimensions of the square torus fixed, i.e., $L=\sqrt{2\pi N_{\phi}}l_B=\sqrt{2\pi N_{\phi}'}l_B'$. Then we have the expression of charge gap used in the main text:
	\begin{eqnarray}
		\Delta_c(N_e)&=&E(N_e,N_{\phi,0}+1)+E(N_e,N_{\phi,0}-1)-2E(N_e,N_{\phi,0})\nonumber\\
		&=&E_g(N_e,N_{\phi,0}+1)+E_g(N_e,N_{\phi,0}-1)-2E_g(N_e,N_{\phi,0})-\frac{2}{9}S(L).
		\label{eq::S5}
	\end{eqnarray}
	We choose the Coulomb energy $E_C$ determined by the magnetic length in the absence of quasiparticles and quasiholes as a unified energy unit. For the energies at other magnetic fields, we obtain their values in units of $E_C$ using 
	\begin{equation}
		E_C'=\frac{e^2}{4\pi\epsilon l_B^\prime}=\sqrt{\frac{N_{\phi}^\prime}{N_{\phi,0}}}E_C,
	\end{equation} 
	where $N_\phi'=N_{\phi,0}\pm 1$. For disordered systems, we average $E_{g}(N_{e},N_{\phi,0}\pm 1)$ and $E_{g}(N_{e},N_{\phi,0})$ over various impurity configurations.

	For systems with finite layer thickness and nonzero distance between the electron and donor layers, $V_{\rm bb},V_{\rm eb},V_{\rm ee}$ contain different $V_{\bm q}$'s. However, the cancellation of the $q= 0$ singularity remains robust. Compared with Eq.~(\ref{eq::S5}), some extra non-singular terms appear due to the layer thickness and separation, but they will vanish in the thermodynamic limit~\cite{WanXin(2005)}. Therefore, we compute the $E_g$'s and $S(L)$ with finite thickness of the electron layer and still use Eq.~(\ref{eq::S5}) for the charge gap. The Madelung energy $S$ for a square lattice is calculated using the Ewald summation method~\cite{Ewald(1917), Fujiki(1992),Bonsall(1977)}:
	\begin{equation}
		\begin{split}
			S(L)=					
			&-\frac{\sqrt{\pi}}{2}\frac{e^2}{4\pi\epsilon L^2}\int_{t_0}^{\infty}dt\ t^{-\frac{3}{2}}f(t)
			-\frac{1}{2\sqrt{\pi}}\frac{e^2}{4\pi\epsilon}\int_0^{t_0}dt\ t^{-\frac{1}{2}}f(t)\\
			&+\frac{1}{2\sqrt{\pi}}\frac{e^2}{4\pi\epsilon}\sum_{\bm{r}({\bm l})\neq0}\tilde{\Theta}(f(t),t_0,|\bm{r}({\bm l})|^2)
			+\frac{\sqrt{\pi}}{2}\frac{\pi e^2}{4\pi\epsilon L^2}\sum_{\bm{G}\neq0}\tilde{\Theta}\left(t^{-1},t_0^{-1},\frac{|\bm{G}|^2}{4}\right),\\
		\end{split}
	\end{equation}
	where 
	\begin{equation}
		\begin{split}
			&f(t)=\int_0^{\infty} dz_1\int_0^{\infty}dz_2|\eta_0(z_1)|^2|\eta_0(z_2)|^2e^{-t|z_1-z_2|^2},\\
			&\tilde{\Theta}(f(t),t_0,x)=\int_{t_0}^{\infty}dt\ t^{-\frac{1}{2}}e^{-2xt}f(t).\\
		\end{split}
	\end{equation}
	Here $\eta_0(z)$ is the lowest subband wave function of electrons in the vertical $z$-direction in main text. The function $f(t)$ encodes the thickness effect of the electron layer by averaging the interaction over the charge distribution in the $z$-direction.  $t_0=\pi/L^2$ is the splitting parameter in the Ewald method, which separates the summation into a short-range part in real space and a long-range part in reciprocal space to ensure rapid convergence~\cite{Ewald(1917),Fujiki(1992)}. In the limit of zero thickness of the electron layer,  we have $S\approx-\frac{3.9}{2}\frac{e^2}{L}$~\cite{Bonsall(1977)}.		
	
		\subsection{C. Many-body Chern number}	
		In this section, we introduce the details of our many-body Chern number calculations. Under the twisted boundary condition, the many-body eigenstates $|\Phi_{m}(\bm{\theta})\rangle$ depend on the boundary phase $\bm{\theta}=(\theta_x,\theta_y)$, with $\theta_x,\theta_y\in [0,2\pi]$. We divide the boundary-phase space into $N_{\theta}\times N_{\theta}$ meshes: 
		\begin{equation}
			(\theta_x,\theta_y)=\frac{2\pi}{N_{ \theta}}(\theta_1,\theta_2),\ \text{with}\ \theta_1,\theta_2=0,1,...,N_{ \theta}-1.
		\end{equation}
		The $n_g$-th group of many-body eigenstates can be expressed as a row vector
		\begin{equation}
			\begin{split}
				\Psi=&\left(|\Phi_{3n_g-2}\rangle,
				|\Phi_{3n_g-1}\rangle,
				|\Phi_{3n_g}\rangle\right).
			\end{split}
		\end{equation}
		Following the method in Ref.~\cite{Fukui(2005)}, we calculate the many-body Chern number for this group under a specific disorder configuration as
		\begin{equation}\label{E_Chern_number}
			C=\frac{1}{2\pi i}\sum_{\theta_1,\theta_2=0}^{N_{\theta}-1}\tilde{F}_{12}(\theta_1,\theta_2),
		\end{equation}
		with 
		\begin{equation}
			\begin{split}
				&\tilde{F}_{12}(\theta_1,\theta_2)=\ln\{U_{1}(\theta_1,\theta_2)U_{2}(\theta_1+1,\theta_2)
				 U^{-1}_{1}(\theta_1,\theta_2+1)U^{-1}_{2}(\theta_1,\theta_2)\},\\
				&U_{1}(\theta_1,\theta_2)=\frac{\text{det}[\Psi^{\dagger}(\theta_1,\theta_2)\Psi(\theta_1+1,\theta_2)]}{|\text{det}[\Psi^{\dagger}(\theta_1,\theta_2)\Psi(\theta_1+1,\theta_2)]|},\
				U_{2}(\theta_1,\theta_2)=\frac{\text{det}[\Psi^{\dagger}(\theta_1,\theta_2)\Psi(\theta_1,\theta_2+1)]}{|\text{det}[\Psi^{\dagger}(\theta_1,\theta_2)\Psi(\theta_1,\theta_2+1)]|}.
			\end{split}
		\end{equation}
		$\Phi(\bm{\theta})$ is obtained by exact diagonalization at each specific boundary condition and the disorder configuration is fixed when the boundary condition changes. We constrain $-\pi<\tilde{F}_{12}(\theta_1,\theta_2)\leq\pi$. We find the many-body Chern number has converged when $N_\theta=10$. We hence choose this value of $N_\theta$ throughout our calculations.

		\subsection{D. An example of identifying the mobility edge}
		
\begin{figure}[htpb]
			\centerline{\includegraphics[width=0.6\linewidth]{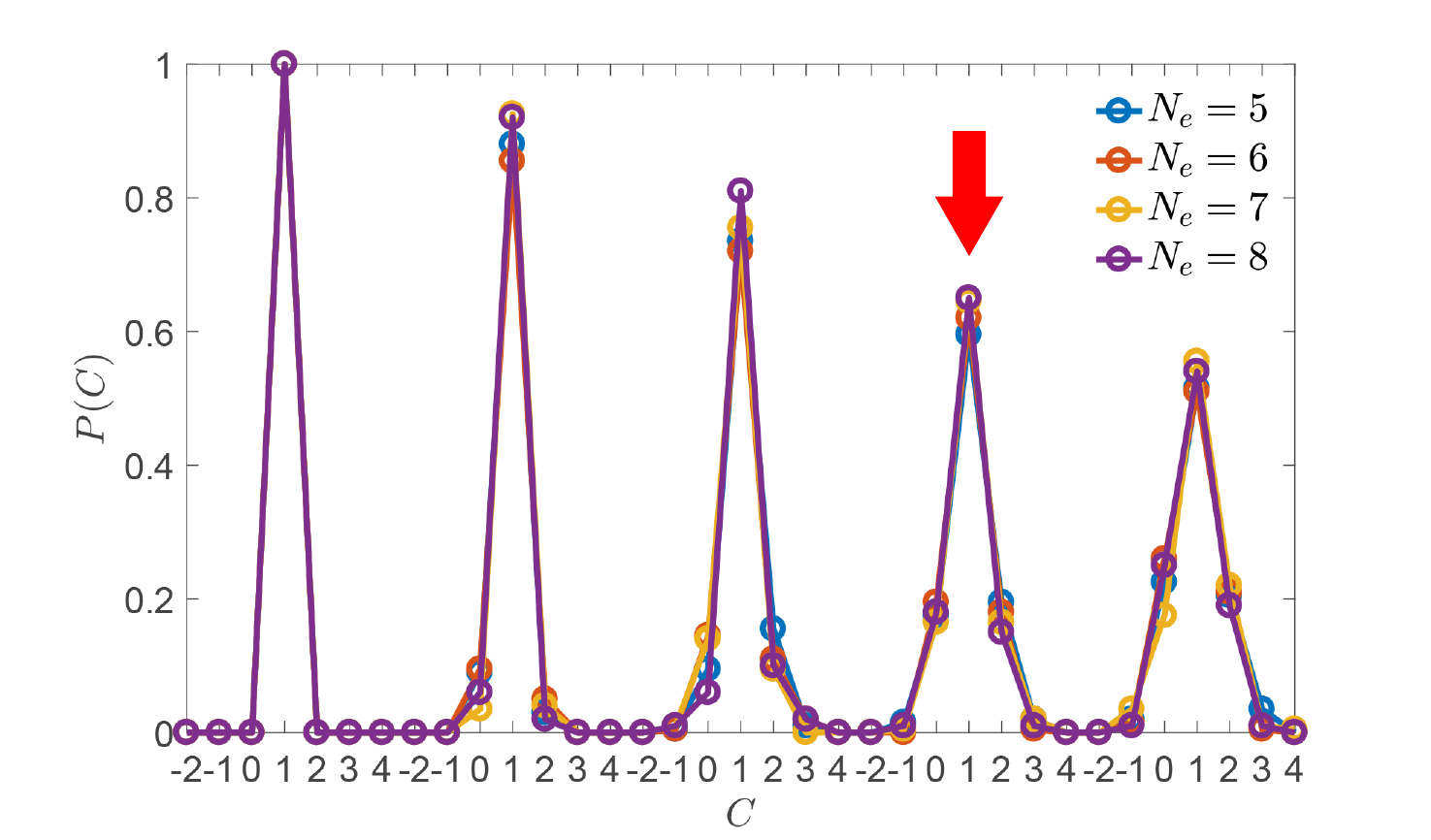}}
			\caption{Probability distribution of Chern number $P(C)$ for the first five groups of states for $N_e=5-8$ at $w=50$ nm. The red arrows indicate the group of states at the mobility edge. We use $200$ disorder configurations for each system size to evaluate $P(C)$.}
			\label{F_Chern_number_example}
		\end{figure}

		As an example, in Fig.~\ref{F_Chern_number_example} we present the probability distribution $P(C)$ of the Chern number $C$ for the lowest five groups of $N_e=5-8$ electrons at $w=50$ nm. For all system sizes, $P(C)$ of the ground-state manifold ($n_g=1$) only has weight at $C=1$, so the Chern number per state is $1/3$, matching the quantized Hall conductance of the $\nu=1/3$ Laughlin state. For groups beyond the ground-state manifold, $P(C)$ starts to develop weights on $C\neq 1$. $P(C=1)$ remains close to $1$ for second and third groups. However, there is the sharpest jump in $P(C=1)$ between third and fourth groups, which is a striking signal of delocalized excitations that can significantly change the transport of the system. Thus we identify it as the mobility edge (red arrow). The mobility gap of a specific system size is estimated as the disorder-averaged difference between the mean energy of the group at the mobility edge and the ground-state group. We emphasize that the group index at the mobility edge may vary with the system size for other values of $w$, especially in weakly disordered systems. 
		
\end{document}